\documentclass{article}
\font\openface=msbm10 at10pt 
\usepackage{graphics}
\def\spose#1{\hbox to 0pt{#1\hss}}
\def\lta{\mathrel{\spose{\lower 3pt\hbox{$\mathchar"218$}}
     \raise 2.0pt\hbox{$\mathchar"13C$}}}
\def\gta{\mathrel{\spose{\lower 3pt\hbox{$\mathchar"218$}}
     \raise 2.0pt\hbox{$\mathchar"13E$}}}
\newcommand{\be}{\begin{equation}}
\newcommand{\en}{\end{equation}}
\newcommand{\bea}{\begin{eqnarray}}
\newcommand{\ena}{\end{eqnarray}}

\newcommand{\Tr}{\mathrm{Tr}}
\newcommand{\lap}{\mathcal{L}^2}
\newcommand{\mat}{\mathrm{Mat}}
\newcommand{\ex}{\mathrm{e}}
\begin{document}
\title{A matrix phase for the $\phi^4$ scalar field on the fuzzy sphere}
\author{XAVIER MARTIN\footnote{xavier@stp.dias.ie} \\ School of
 Theoretical Physics, DIAS, 10 Burlington Road, Dublin 4, IRELAND} 
\maketitle
\begin{abstract}
The critical properties of the real $\phi^4$ scalar field theory are
studied numerically on the fuzzy sphere. The fuzzy sphere is a finite
matrix (non--commutative) approximation of the algebra of functions on
the usual two dimensional sphere. It is also one of the simplest
examples of a non--commutative space to study field theory on.  Aside
from the usual disordered and uniform phases present in the
commutative scalar field theory, we find and discuss in detail a new
phase called a matrix phase because the geometry of the fuzzy sphere,
as expressed by the kinetic term, becomes negligible there. This
highlights a new aspect of $UV$--$IR$ mixing, the unusual behaviour
which arises naturally when taking the commutative limit of a non
commutative field theory.
\end{abstract}

\section{Introduction}
The fuzzy approximation scheme \cite{fuzzy,badis1} consists in
approximating the algebra of functions on a manifold with a finite
dimensional algebra (i.e. a matrix algebra) instead of discretising
the underlying space as a lattice approximation does.

Studying field theory on a non--commutative space such as the matrix
algebras we are considering is covered by the framework of Connes'
non-commutative geometry \cite{Connes}. For reasons of simplicity,
only the simpler case of the real scalar field will be considered in
this paper. As a result, the only relevant differential operator is
the Laplacian and it is possible to consider a greatly simplified theory.

Indeed, approximating an algebra of functions as an algebra of
matrices does not carry in itself any geometrical content, as is
obvious from the fact that all fuzzy approximations must yield the
same matrix algebras. What really defines the geometry is the choice
of a derivation and scalar product. For the scalar product, we will
normally arrange to use the canonical scalar product on the algebra of
matrices \be <\phi|\psi> \propto \Tr(\phi^* \psi).\label{sp} \en 
As a consequence, complex conjugation will be associated with
hermitian conjugation and thus, real fields will be approximated with
hermitian matrices. Then, as far as the scalar field theory is
concerned, the only required differential operator is the
Laplacian. The fuzzy approximation of the scalar field theory on a
manifold will therefore be entirely determined by the choice of a
Laplacian and multiplicative coefficient in (\ref{sp}) for each
matrix algebra.

However, even with such a simple scheme, only a few manifolds can be
``fuzzified'' in this way, including the complex projective planes
$\mbox{\openface CP}^N$ \cite{BDLMO} as well as their Cartesian
products \cite{badis2}. Under certain conditions, it is also possible
to approximate other spaces such as the spheres $S^3$ \cite{DB} and
$S^4$ \cite{DJ} by imbedding them in one of the fuzzy $\mbox{\openface
CP}^N$. The non--commutative lattice \cite{ftorus} is another simple
matrix space where numerical simulations have already been performed,
although it is not a fuzzy space in the sense that it does not
approximate a manifold but a lattice.

As a approximation scheme, this ``fuzzification'' is well suited to
numerical simulations of field theories \cite{Nishi}. As a test run,
the first fuzzy approximation to be investigated should be the
simplest one, that of the two dimensional sphere $\mbox{\openface
CP}^1=S^2$.  Besides, the two--dimensional plane can be viewed as the
limit of a sphere of infinite radius.

In this paper, the fuzzy sphere and its properties will first be
introduced in Section \ref{fuzsph}. Then, the real scalar field theory
on a sphere and on the fuzzy spheres will be presented in Section
\ref{RSFT}. The following section \ref{simuls} shows the results of
the simulations. The phase boundaries between the three phases present
and their scaling properties are derived. A careful analysis of this
new phase is also presented in this section.  Finally, the results are
summarised and discussed in the Conclusion \ref{Conclu}.

\section{The fuzzy sphere}\label{fuzsph}
The simplest example of a fuzzy space is the fuzzy sphere
\cite{Madore}. As explained in the Introduction, for the purpose of
studying a scalar field theory, the only ingredient required to fix
the geometry is a Laplacian operator and a scalar product on each
matrix algebra. Since derivations on the commutative sphere can be
viewed as infinitesimal $S\mathrm{SU}(2)$ transformations, the Laplacian on a
$(2s+1)\times (2s+1)$ matrix algebra, also denoted $\mat_{2s+1}$, can
be guessed as \be
\lap \phi=[L_i,[L_i,\phi]], \label{flap} \en 
where $L_i$ are the angular momentum operators in the $2s+1$
dimensional irreducible representation of $\mathrm{SU}(2)$. The scalar product
is chosen as proposed in (\ref{sp}), with a multiplicative coefficient
such that the unit matrix has the same norm as the unit function on
the sphere \be
<\phi|\psi>=\frac{4\pi}{2s+1} \Tr(\phi^\dag \psi). \label{scpr} \en

The spectrum of the proposed Laplacian operator can be recognised
from the adjoint action of angular momentum as \be
\lap \hat{Y}_{lm}=l(l+1)\hat{Y}_{lm} ,\ 0\leq l\leq 2s, \label{fsh} \en
where the eigenfunctions $\hat{Y}_{lm}$ are the polarisation tensors
 whose normalisation is defined according to the chosen scalar product \be
\frac{4\pi}{2s+1} \Tr(\hat{Y}_{lm}^\dag \hat{Y}_{lm})=1.\en
This is precisely the spectrum of the Laplacian on the commutative
sphere truncated at angular momentum $2s$, thus vindicating this
choice.

A clean way of recognising the approximation of a sphere in these
matrix algebras is to introduce a mapping which associates a function
on the sphere with each matrix of the algebra $\mat_{2s+1}$ and pulls
back most of the structure on the algebra of functions of the sphere
onto the matrix algebra \cite{denjoemap}. There are various ways to
define such a mapping, such as using coherent states
\cite{pres}. However, the simplest one is given by \bea
\mathcal{M}_s: \mat_{2s+1} & \rightarrow & \mathcal{C}^\infty (S^2) \\
M=\sum_{l=0}^{2s}\sum_{m=-l}^l c_{lm}\hat{Y}_{lm} & \mapsto & 
f({\bf n})=\sum_{l=0}^{2s}\sum_{m=-l}^l c_{lm} Y_{lm}({\bf n}),
\label{map} \ena
where the functions $Y_{lm}({\bf n})$ are the usual spherical harmonics
on the sphere, i.e. the eigenvectors of the Laplacian operator on the
sphere. By definition, this mapping $\mathcal{M}_s$ is linear and maps
the Laplacian $\lap$ on $\mat_{2s+1}$ onto the Laplacian on the
sphere. In fact, the three
derivatives on the sphere $\nabla_l=i\varepsilon _{jkl} x_j
\partial_k$ are pulled back to simple derivations on the matrix
algebra given by \be
\mathcal{L}_i\phi=[L_i,\phi ] .\en
By construction, the action of the group $\mathrm{SU}(2)$ is preserved on both
sides. Furthermore, since the eigenvectors of the Laplacian on the
matrix space and on the sphere form orthonormal bases on their
respective spaces, this mapping is an injective isometry. Its image,
on which the mapping is one to one, $\mathcal{M}_s(\mat_{2s+1})$ is
given by all the functions with angular momentum only up to $2s$ and
form a sequence of increasing (for the inclusion) sets which become
dense in $\mathcal{C}^\infty (S^2)$ in the limit of infinite
matrices. The matrix product is mapped to a (non--commutative) product
of functions on the sphere called a $*$--product \be
\mathcal{M}_s (\phi\psi)({\bf n})=(\mathcal{M}_s(\phi)*_s
\mathcal{M}_s(\psi)) ({\bf n}),\en
which is evidently distinct from the usual (commutative) product of
functions. It is possible to verify that in the limit of infinite
matrices $s\rightarrow \infty$, the star product tends to the usual
product. More precisely, for $(f_s,g_s)\in (\mathcal{M}_s(\mat
_{2s+1}))^2$ two functions with angular momentum truncated at $2s$,
and $t\geq s$, \be 
(f_s*_t g_s)({\bf n})=f_s({\bf n})g_s({\bf n})+\mathcal{O}(
\frac{1}{t}).\en

Note in passing that complex conjugation of a function on the sphere
pulls back to hermitian conjugation on the matrix algebra.
Consequently, as proposed in the introduction, real functions pull
back to hermitian matrices. Similarly, integration on the sphere
which is similar to scalar product with the unit function pulls back
to the trace on the matrix algebra.

Thus, in the limit when $s$ goes to infinity, the mapping
$\mathcal{M}_s$ becomes an isomorphism of algebras which preserves
rotational invariance, the Laplacian and the scalar product
(\ref{scpr}). This proves that the fuzzy spaces, as defined by the
triple $(\mat_{2s+1},\lap,<\cdot,\cdot>)$ go over to the sphere in
the limit of infinitely large matrices. Furthermore, we can deduce
immediately the following approximation rule.

{\em Approximation rule:} given an algebraic expression on the
sphere, it is possible to find a fuzzy approximation for it, which
converges to it in the large matrix limit, by truncating all functions
at momentum $2s$, replacing products by $*_s$--products everywhere and
pulling back the expression into the matrix algebras with the mapping
$\mathcal{M}_s$.

Another mapping with similar properties which is generally introduced
is the one obtained by looking at the diagonal elements of a matrix in
a coherent states representation. Compared to $\mathcal{M}_s$, this
mapping trades the isometry property for the conservation of the
notion of state, in the sense that it maps a state of $\mat_{2s+1}$
into a state of $\mathcal{C}^\infty (S^2)$. In this case, the
corresponding star product can also be expressed in a simple exact
form \cite{pres}.

This introduction to the fuzzy sphere described spheres of radius
one. Getting spheres of different radius $R$, is just a matter of
scaling the scalar product (\ref{sp}) and Laplacian (\ref{flap})
appropriately: \bea 
\lap & \rightarrow & \frac{1}{R^2} \lap ,\label{scaleL} \\ 
\frac{4\pi}{2s+1} \Tr(\phi^\dag \psi) & \rightarrow & \frac{4\pi
  R^2}{2s+1} \Tr(\phi^\dag \psi) .\label{scaletr} \ena 

With the fuzzy sphere now defined and understood, we can move on to
defining a real scalar field theory on it.

\section{The real scalar field theory} \label{RSFT}
The $\phi^4$ scalar field theory on the two--dimensional sphere is
given by the action\footnote{A more usual form of this action can be
  obtained by changing the field $\phi$ to $\phi/\sqrt{2}$. Since the
  observables evaluated in this paper are typically homogeneous of
  degree two in the field, they are simply multiplied by an overall
  factor of two in this change of variable.} \be
S(\phi)=\int_{S^2} d^2{\bf n} (\phi\Delta \phi +r\phi^2+\lambda
\phi^4), \label{scalact} \en 
with $\phi$ a real scalar field, $\Delta=-\nabla_i\nabla_i$ the
Laplacian on the sphere, $r$ a mass parameter and $\lambda$ an
interaction constant. This particular model was chosen because it is
simple and well studied. In fact, it is known that the diagrammatic
expansion of this theory has only one divergent diagram, the tadpole
diagram, is Borel resumable, and defines the field theory entirely.

Using the Approximation rule above, the action (\ref{RSFT}) for the
real scalar field can be approximated by \cite{den} \be
S(\phi)=\frac{4\pi}{2s+1}\Tr (\phi\lap \phi +r\phi^2+\lambda
\phi^4),\label{Fscalact} \en 
where $\phi$ must be an hermitian matrix. Again, it is possible to
write a diagrammatic expansion for this theory \cite{den}. There are
more diagrams than in the continuum since the legs of the vertices do
not commute anymore, although they can still be cyclically rotated.
On the other hand, since the theory is defined on a finite dimensional
algebra, all diagrams must be finite. Furthermore, since the action
was obtained by the approximation rule, it is easy to check that all
the fuzzy diagrams are just approximations of their commutative
counterparts obtained from the action (\ref{RSFT}). Thus, in the limit
of infinite matrices, all finite diagrams converge to their commutative
counterparts, while the tadpole diagrams must also diverge.

The approximation rule says nothing of the sub-dominant contribution
to the tadpole diagrams however. In fact, the constant contribution of
one of the two diagrams (the ``non--planar'' one where the exterior
legs are not adjacent) does {\em not} converge to its usual continuum
limit \cite{den}. Thus, we see that the field theory described by the
action (\ref{Fscalact}) is {\em not} an approximation of the continuum
field theory (\ref{scalact}). This is what gives rise to the so called
$UV$--$IR$ mixing in the disordered phase, as studied in
\cite{den}. In the following, I will generically refer to ``theories
with $UV$--$IR$ mixing'' as such theories which do not converge to
their commutative limits.

At this point, there are two interesting theories which can be
studied. The first one described by the fuzzy action (\ref{Fscalact})
with its associated $UV$--$IR$ mixing is a simple example of a
non--commutative theory such as those that have cropped up in particle
theory recently. The other one is described by the same action
(\ref{Fscalact}) plus an additional damping term to ensure that the
fuzzy theory does indeed go over to the commutative one
\cite{den}. Such a fuzzy theory will then be an alternative to the
usual lattice discretisation of a scalar field theory.

As the first step toward a possible new approximation method of field
theory, the latter theory holds the most potential. However, the
former theory, which will be studied in this paper, proposes a
non--perturbative analysis of a non--commutative field theory with
$UV$--$IR$ mixing, and more prosaically is simpler to implement. A
similar study was done for this same model on the non--commutative
lattice in \cite{ftorus}.

Even though the scalar field on the commutative or fuzzy spheres
cannot have a phase transition since both have finite volume, one may
be found in the planar limit, i.e. in the limit where the matrices and
the sphere become infinite $s, R\rightarrow\infty$. It is therefore
convenient in the following to introduce explicitly the sphere radius
$R$ in the problem.

From Eqs. (\ref{scaleL},\ref{scaletr}), it is clear that the real
scalar field action with variable radius must take the form \be
S(\phi)=\frac{4\pi}{2s+1}\Tr (\phi\lap \phi +rR^2\phi^2+\lambda
R^2\phi^4).\label{FscalactR} \en 
It will also be convenient later on to define the potential part
of the action given by \be
V(\phi)=\frac{4\pi R^2}{2s+1}\Tr (r\phi^2+\lambda \phi^4).\label{potV} \en 
Of the three parameters $r$, $\lambda$ and $R$, only two are
independent, and in the following, according to the situation, either the
interaction parameter $\lambda$ or the radius $R$ will be set to one
while the other parameter is allowed to vary.

\section{The simulations} \label{simuls}
The theory simulated here is the one described by the action
(\ref{FscalactR}) with fields $\phi$ which are hermitian matrices. At
least one critical line is expected to arise when the radius $R$ goes
to infinity, corresponding to the critical behaviour of the scalar
field on the plane. Thus, the goal of the simulation will
be to draw a phase diagram for this theory.

The simplest way to test for a continuous phase transition is to look
for peaks of the finite--volume susceptibility\footnote{The usual
  susceptibility has no peak for finite volume because by symmetry
  $<\Tr(\phi)>=0$. Instead the standard finite--volume susceptibility
  is used where the latter term is replaced by $<|\Tr(\phi)|>$ which
  has the right properties.} given by \be
\chi=<\Tr(\phi)^2>-<|\Tr(\phi)|>^2, \label{susc} \en
where the expectation values are given by \be
<h(\phi)>=\int \mathrm{d}\phi \, \frac{\ex^{-S(\phi)}}{Z}
h(\phi),\label{ev} \en 
with $S$ the field action (\ref{FscalactR}), $h$ some algebraic
expression in the field, and $Z$ the partition function \be 
Z=\int \mathrm{d}\phi \, \ex^{-S(\phi)} .\en
For our purpose, the normalisation of the susceptibility is
unimportant, contrary to the observables introduced next.

To get a better idea of the phase the field is in, it is also
convenient to monitor a few other significant quantities. Using the
mapping $\mathcal{M}$ introduced in (\ref{map}), it is possible to associate
functions on the sphere to matrices. Therefore, the coefficients
$c_{lm}$ in the expansion of a matrix on the basis of polarisation
tensors (\ref{fsh}) \be 
\phi=\sum_{l,m}c_{lm}\hat{Y}_{lm},\en
have an immediate classical interpretation. They will automatically
average to zero though since they are linear, and thus odd, in the
field while the action is even in it. For this reason, we chose
the following quadratic even additional observables \bea 
<\| \phi\| ^2> & = & <\sum_{l,m} |c_{lm}|^2>=<\frac{4\pi}{2s+1}
\Tr(\phi^2)> \label{pow} \\ 
<|c_{00}|^2> & = & \frac{4\pi}{(2s+1)^2}<(\Tr(\phi))^2>. \label{pow0} \\
<\sum_{m=-1}^1 |c_{1m}|^2> & = & \frac{12\pi}{s(s+1)(2s+1)^2} [ \,
|\Tr(\phi L_+)|^2+(\Tr(\phi L_0))^2 ] ,\label{pow1} \ena
where $L_+=L_1+iL_2\propto \hat{Y}_{11}$. The first two observables also
happen to be invariant under $\mathrm{U}(n)$ transformations while the latter
is only invariant under $\mathrm{SU}(2)$ transformations. These quantities can
respectively be related to the average of the total power of the
field, and of its power in the modes $0$, and $1$.

The simulations themselves are realised in the standard way using a
Metropolis Monte-Carlo method with the jackknife method to evaluate the
error on the calculated expected values \cite{LB}. To better control
thermalisation, simulations were run starting from both hot
(i.e. random) and cold (i.e. the minimum of the action) initial
conditions.

This generates a sequence of random field configurations $\phi(t)$
with probability distribution given by $\ex^{-S}/Z$ as found in the
expectation values (\ref{ev}). The parameter $t$ will be called
``Monte--Carlo time'', and the expectation values will then be
calculated as standard averages \be
<h(\phi)>=\frac{1}{T}\sum_{t=1}^T h(\phi(t)).\label{av} \en
The update of the matrix $\phi (t)$ through one Monte--Carlo time step
is done entry by entry as this is not slower than a global matrix
change. With the acceptance rate defined as the number of entries
modifed during a sweep over the total number of entries tested (which
is all of them), the range of variation of each matrix entry is fixed
adaptatively by maintaining the acceptance rate between $15\%$ and
$30\%$.

The key model--dependent ingredient in this method is the calculation
of the variation of the action under a random variation of a matrix
entry, $\delta S_{ij}(x)=S(\phi_{ij}\rightarrow \phi_{ij}+x)$ which is
used to calculate the probability $\min(\ex^{-\delta S},1)$ of
accepting $\phi_{ij}+x$ as the matrix entry at the next Monte--Carlo
timestep, and the calculation of the observables which is used in the
average formula (\ref{av}) and is always negligible in terms of the
number of operations compared to $\delta S_{ij}(x)$.

The scalar field on the fuzzy sphere is ``non--local'' in the sense
that a matrix entry $\phi_{ij}$ is coupled to all the other entries on
its line $\phi_{ik}$ and column $\phi_{kj}$ through the interaction
term of the potential $\lambda\phi^4$. As a result, the variation of
the action $\delta S$ requires a number of operations which grows
linearly with the matrix size $2s+1$. Since each entry of the matrix
must be updated from one Monte--Carlo timestep to the next, the
computation time of a matrix update for the fuzzy action
(\ref{FscalactR}) grows like $\mathcal{O}(s^3)$. The quadratic part of
the action only couples it to a fixed number of other entries, so that
$\phi_{ij}$ only couples to $\phi_{i+1\, j+1}$, $\phi_{ij}$ and
$\phi_{i-1\, j-1}$, and is therefore subdominant in this calculation.

By comparison, for a ``local'' action, such as that of a finite
difference scalar field action on the lattice, a degree of freedom (or
lattice site) is only coupled to a constant number of other degrees of
freedom, i.e. independant of the total number of degrees of freedom,
and the number of operations to update the field through one
Monte--Carlo timestep only grows like the number of degrees of freedom
$\mathcal{O}(s^2)$.

This calls for two observations. First, the only ``non--local term''
in the fuzzy action is actually the self--interaction term $\lambda
\phi^4$. It should be noted in passing that the non--locality gets
worse as the power of the self--interacting term increases, such as
adding a term in $\Tr(\phi^6)$, in the sense that the number of
operations to calculate $\delta S_{ij}$ grows polynomially even
faster. Second, although this model cannot be compared to a lattice
model because of the $UV$--$IR$ mixing, a corrected action such as those
proposed in \cite{den} which would converge towards the clasical real
scalar field theory would seem to be intrinsically slower than its
lattice equivalent. This however does not generalise to other field
theories, particularly fermionic ones, and does not take into account
the rate of convergence of the Monte--Carlo scheme itself.

Let us now go over the various phases which arise in the
simulation. In the following, as a convention when either of the two
parameters $\lambda$ or $R$ is not mentionned as a variable, it is
assumed to be equal to one.

\subsection{The uniform--disordered phase transition} \label{disuni}
\begin{figure} \input{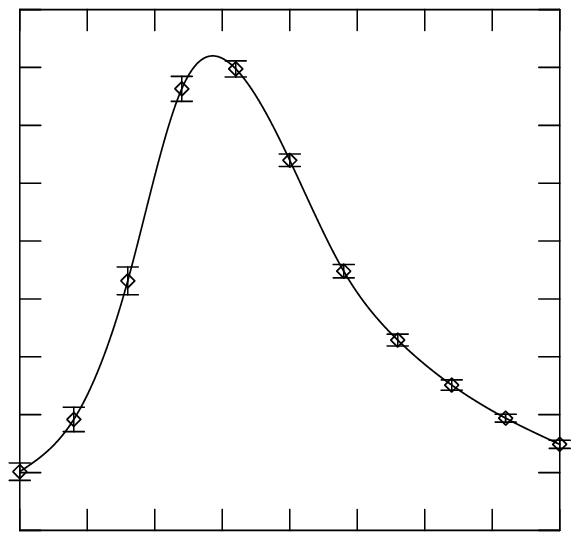} \input{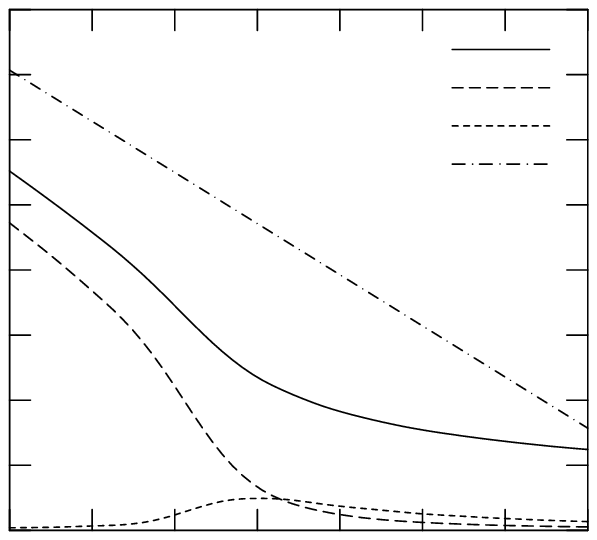}
\caption{ (a) Plot of the susceptibility from Eq. (\ref{susc}),
  calculated on $2^{16}$ bins of $40$ Monte--Carlo timesteps for
  $\lambda=8$ and $21\times 21$ matrices. Its maximum is evaluated at
  $r^{u-d}_c(\lambda=8)=-13.2\pm 0.6$. (b) Plot of the three
  observables described in Eqs.  (\ref{pow},\ref{pow0},\ref{pow1}) for
  the same parameters.}
\label{ord-dis} \end{figure}

This is the phase transition observed for a $\phi^4$ scalar field
theory on the commutative plane. For the model considered here, it
appears for small interaction parameters $\lambda$. Figure
\ref{ord-dis} shows a typical example of this phase transition. The
critical point can readily be identified as the maximum of the
susceptibility (\ref{susc}).

As seen on Figure \ref{ord-dis}b, the observables
above the critical point $r>r^{u-d}_c$ show no strong dependance on
$r$. The power in the $0$ and $1$ modes is also much smaller than the
total power suggesting that the random fluctuations are spread out
over all the modes. This is characteristic of a disordered phase where
the field is composed entirely of random fluctuations around the
constant average field $\phi=0$. This phase is found when the mass
parameter $r$ is positive or negative and ``small'', $-r\lta
\lambda$. These results are completely consistent with what one would
expect from a commutative scalar field theory.

The uniform phase arises when the mass parameter $r$ is negative and
``large''. It appears in Figure \ref{ord-dis} below the critical point
$r<r^{u-d}_c$. In this approximation, the action is completely dominated
by the regions around its minima at $\phi_\pm=\pm\sqrt{-r/2\lambda}\,
{\bf 1}$, where ${\bf 1}$ denotes the unit matrix. In this limit, it
is possible to expand the action to dominant (quadratic) order around
these two minima to derive the observables. This gives an effective
action which is basically the sum of two delta distributions centered
at each of the action minima $\phi_\pm$. In particular, as seen in
Figure \ref{ord-dis}b, \bea
<\| \phi \| ^2> \simeq <|c_{00}|^2> \simeq \frac{-2\pi r}{\lambda} \\
<\sum_m|c_{1m}|^2> \ll \ <|c_{00}|^2> .\ena 
Checking the next order of expansion shows that the power in all the
non--zero modes is suppressed as $1/(-r)$.
\begin{figure} 
\input{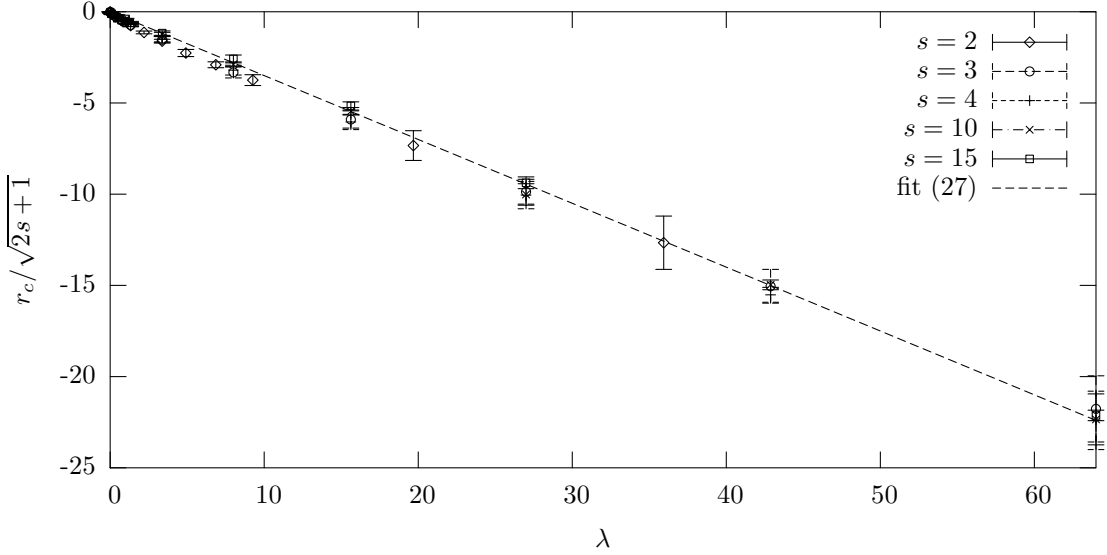}
\caption{Scaled phase diagram for the uniform--disordered phase
 transition for five distinct matrix sizes $(2s+1)\times(2s+1)$ and the
 best fit given by (\protect\ref{fit1}).} \label{phase1} \end{figure}

The corresponding critical line in the phase diagram is shown in
Figure \ref{phase1}. The parameters $r$ and $\lambda$ were chosen as
variables because the critical line goes through the point $(0,0)$
around which it is delicate to scale the parameters using the radius
$R^2$. This critical line scales like the square root of the matrix
size. Furthermore, a simple fit of the datas suggests that it is well
approximated by \be 
\frac{r^{u-d}_c(\lambda)}{\sqrt{2s+1}} \simeq -0.35\lambda.\label{fit1} \en 
The phase boundary line is well approximated by a linear function, but
its slope seems to change slightly with the matrix size. The fit
proposed above has been done on the largest matrix size data, $s=15$,
which is hopefully the best approximation to the large $s$ limit.

It would be interesting to study in more detail this phase transition
and how it compares to the disordered--uniform phase transition of the
commutative scalar field theory near the origin. However, the focus of
this paper is the new phase, which is found and described in the next
subsection. This study is therefore left for a future paper \cite{fergar}.

\subsection{The disordered--matrix phase transition}\label{dmphase}
\begin{figure} \input{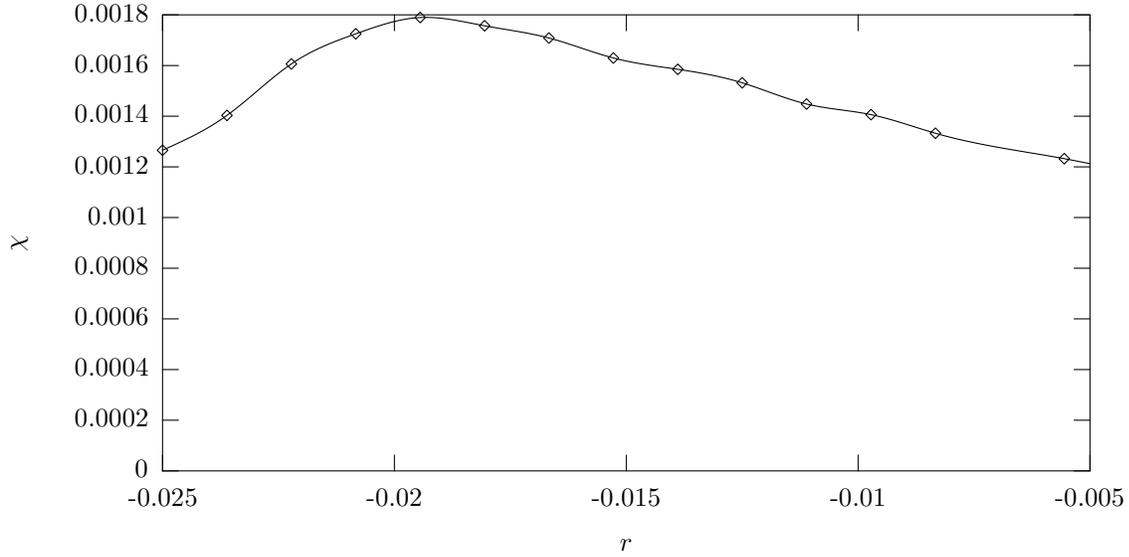} 
\caption{Plot of the susceptibility (\ref{susc}) calculated for
 $21\times 21$ matrices on $2^{16}$ bins of $40$ Monte--Carlo
 timesteps for $R^2=72^3$. Its maximum is evaluated at
 $r^{d-m}_c(72^3)=-0.019\pm 0.005$.}
 \label{dis-mat-chi}\end{figure}

\begin{figure} \input{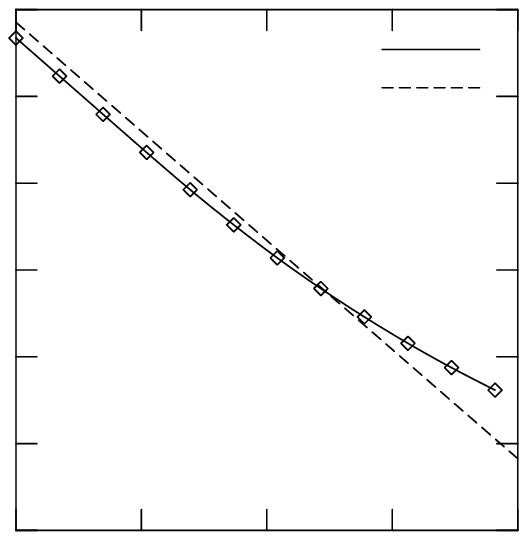} \input{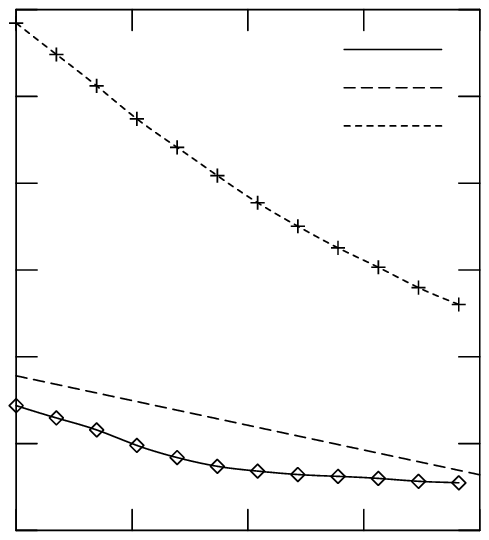}
\caption{(a) Plots of the total power (\ref{pow}) and its proposed
  approximation (\ref{tr2dismat}) calculated for the same parameters
  as in Figure \protect\ref{dis-mat-chi}. (b) Plots of the power in
  the $0$ mode (\ref{pow0}), its proposed approximation
  (\ref{trdismat}), and the power in the $1$ mode (\ref{pow1})
  calculated with the same parameters.}
\label{dis-mat-pwer} \end{figure}

The simulations show the appearance of a new phase for larger radii
which I will call the ``matrix phase'' for reasons explained in the
paragraph ``The pure potential model'' of this subsection. Figures
\ref{dis-mat-chi} and \ref{dis-mat-pwer} show a typical example of
this phase transition. Again the phase transition can be clearly
identified as occuring at the maximum of the susceptibility.

The new phase seems characterised by \bea
<\frac{1}{2s+1}\Tr(\phi^2)> & \simeq & \frac{-2\pi r}{\lambda}
\label{tr2dismat} \\
<c_{00}^2> & \simeq & \frac{-2\pi r}{\lambda(2s+1)^2} \, \ll
\, <\frac{1}{2s+1}\Tr(\phi^2)> \label{trdismat} \\
<\sum_m |c_{1m}|^2 > & > & <c_{00}^2> .\label{approxdismat} \ena
This suggests strongly that in this new phase the field $\phi$ takes
the form $\phi \sim \pm \sqrt{-r/2\lambda} \, U^\dag ({\bf 1}_{s}\oplus
-{\bf 1}_{s+1} )U$ with $U\in \mathrm{U}(2s+1)$ and ${\bf 1}_n$ the $n\times
n$ unit matrix, which is a minimum of the potential (\ref{potV}).

Further examination of the raw data of a Monte--Carlo run with cold
initial conditions shows that during thermalisation $\Tr(\phi(t))$
($t$ being the Monte--Carlo time) goes through a series of plateaus at
$(2l+1)\sqrt{-r/2\lambda}$, $|l|\leq s$ before settling down at its
equilibrium value corresponding to $l=0$, which is consitent with the
value of $<c_{00}^2>$ given in Eq. (\ref{trdismat}). This seems to
confirm that the minima of the potential part of the action given by
$\sqrt{-r/2\lambda} \, U^\dag ({\bf 1}_l \oplus -{\bf 1}_{2s+1-l} )U$
with $U\in \mathrm{U}(2s+1)$ are local minima of the action.

Normally in the uniform phase, the minimum corresponding to $l\in \{
0,2s+1\}$ is automatically selected by virtue of also minimizing the
kinetic part of the action. The fact that in the matrix phase it is
not suggests that the kinetic term might be negligible. To test this
assertion, the field theory described by the action $V(\phi)$ from
Eq. (\ref{potV}) has been studied and compared to the results for the
fuzzy scalar field. 
\paragraph{The pure potential model} This model is actually much
simpler than the fuzzy scalar field action by virtue of being
invariant under $\mathrm{U}(2s+1)$ transformations of the field $\phi
\rightarrow U^\dag \phi U$, $U\in \mathrm{U}(2s+1)$.

Thus, considering only observables which are also invariant under
these transformations, such as those defined in
Eqs. (\ref{susc},\ref{pow},\ref{pow0}), the degrees of freedom
associated with this invariance can be extracted in the form
$\phi=U^\dag \mathrm{Diag}(x_1, \ldots,x_{2s+1}) U$, which has a
Jacobian given by the square of the Vandermonde determinant where the
Vandemonde determinant is given by $\Delta (x_1,\ldots,x_{2s+1})=\prod
_{i<j}(x_i-x_j)$, and integrated out to get an effective action given
by \be V_\mathit{eff}(x_1,\ldots,x_{2s+1})=\frac{4\pi R^2}{2s+1}
\sum_{i=1}^{2s+1} (rx_i^2+\lambda x_i^4)-\sum_{1\leq i<j\leq 2s+1} 
\ln [(x_j-x_i)^2] .\label{Veff} \en
Furthermore, in the case of the observables defined in Eqs.
(\ref{susc},\ref{pow},\ref{pow0}), generically called
$F(r,\lambda,R)$, which are quadratic in the field, a simple change of
variable $y_i=R^{1/2}\lambda^{1/4}$, shows that these averages
effectively depend only on one parameter \be
R\sqrt{\lambda}\, F(r,\lambda,R)=F(r/(R\sqrt\lambda),1,1). \label{scaleV} \en

\begin{figure} \input{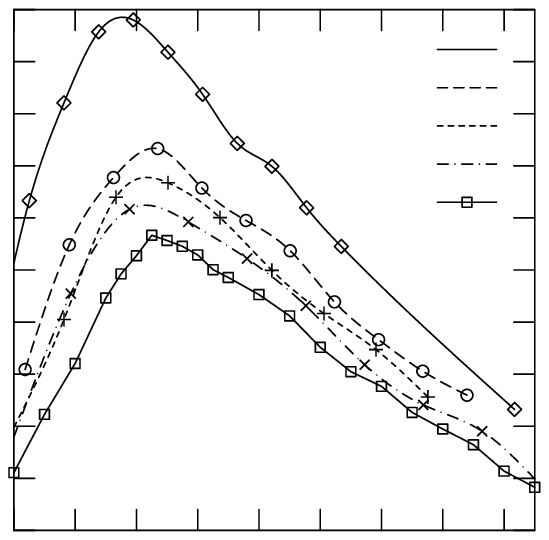} \input{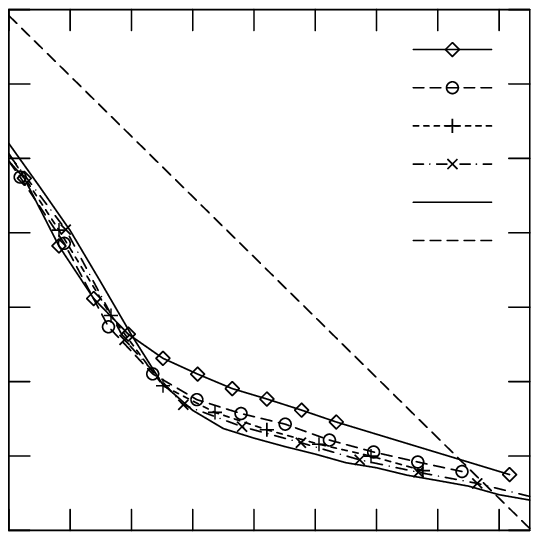}
\caption{(a) Plot of the susceptibility for $31\times 31$ matrices, in
 units such that they do not depend on the radius in the case of the
 pure potential action as shown in Eq. (\ref{scaleV}). Note that the
 critical point for the pure potential model is at $r^{d-m}_c
 R=-17.5\pm 0.5$.  (b) Plot of $<c_{00}^2>$ and its approximation
 (\ref{trdismat}) for the same parameters. In both figures (a) and
 (b), the observables calculated from the fuzzy scalar field action
 seem to converge monotonically with $R^2$ toward the pure potential
 observables.}
 \label{compareff} \end{figure}

This effective action (\ref{Veff}) can be easily studied by
Monte--Carlo simulations since it has a lot fewer degrees of
freedom. Figure \ref{compareff} supports the assertion that around the
phase transition considered here, the kinetic term of the fuzzy scalar
field model becomes negligible in the limit when the radius $R^2$
tends to infinity. This is why the new phase was called a ``matrix
phase''. In this phase, the kinetic term which carries the geometrical
content of the fuzzy sphere vanishes. It is likely that the ``striped
phase'' which has been found for the $\phi^4$ theory on the
non--commutative lattice \cite{ftorus} arises for the same reason and
must therefore converge to the same matrix phase. In particular, for
the same parameters and deep in the matrix phase, the
values of $\mathrm{U}(2s+1)$ invariant observables, such as $<\Tr(\phi)>$,
$<\Tr^2(\phi)>$, or $<\Tr(\phi^2)>$, should be the same on the
non--commutative lattice as on the fuzzy sphere.
\paragraph{The matrix phase} Eq. (\ref{tr2dismat}) suggests that in
the matrix phase the eigenvalues of the field settle in a minimum of
the potential. This suggests that only the neighbourhood of the minima
of the potential contribute to the expectation values (\ref{ev}). 

The minima of the potential are $2s+1$ disjoint orbits of the
form \be
O_n=\{ \sqrt\frac{-r}{2\lambda} U^\dag ({\bf 1}_n\oplus -{\bf
 1}_{2s+1-n}) U \, | U\in \mathrm{U}(2s+1)/ [ \mathrm{U}(n)\times \mathrm{U}(2s+1-n) ] \} ,\en 
where $n\leq s+1/2$ and ${\bf 1}_n$ are the $n\times n$ unit
matrices. These orbits are isomorphic to Grassmanians
$\mathrm{G}^{2s+1} _n \cong \mathrm{U}(2s+1)/[ \mathrm{U}(n)\times \mathrm{U}(2s+1-n)
]$. When the kinetic term is negligible, the field distribution will
be dominated by the orbit with the largest phase space volume which
corresponds to $n$ the integer value out of $s$ and $s+1/2$.

By comparison, the kinetic term which was neglected will select the
diagonal orbits, $n\in \{ 0,2s+1\}$, which have the lowest phase space
volume. Thus, the matrix phase can be interpreted as a phase where the
kinetic term is negligible with respect to the volume of the largest
orbit $\mathcal{O}_s$. 

This reasoning can be verified more rigorously, by looking carefully at
the probability measure \be
\mathrm{d}\mu(\phi)=\frac{\ex^{-S(\phi)}}{Z} {d}\phi \en
asociated with the field theory. The simulations indicated clearly
that only the fields near the minimum of the potential contribute to
the distribution. Therefore, expanding $\phi$ around these minima in
the form \be 
\phi=x_0 \, U^\dag \mathrm{Diag}(\epsilon_j+x_j) \, U,\label{expanphi} \en
with $U\in \mathrm{U}(2s+1)$, $x_0=\sqrt{-r/2\lambda}$ the minimum of the
potential, and $\epsilon_j\in \{-1,+1\}$, the probability distribution
becomes \be 
\mathrm{d}\mu (\phi)\approx \frac{1}{Z}\sum_{\epsilon_j}\delta(\phi
-x_0 \, U^\dag \mathrm{Diag}(\epsilon_j+x_j) \, U)\, \ex^{-4\pi(rR)^2
{\bf x}\cdot {\bf x}/[\lambda(2s+1)]-x_0^2 K(U,\epsilon_j,x_j)} \, V^2(\epsilon_j+x_j)
\,\mathrm{d}U\, \mathrm{d}^{2s+1}{\bf x}, \en 
where \be
K(U,\epsilon_j,x_j)=\frac{4\pi}{2s+1}\Tr [(U^\dag \mathrm{Diag}(\epsilon
_j+x_j) \, U) \lap (U^\dag \mathrm{Diag}(\epsilon_j+x_j) \, U)] \en 
is the kinetic term for the field expanded acording to
Eq. (\ref{expanphi}) and $Z$ is the partition function whose value
changes as needed to normalise the probability distribution it is part
of. It is possible to simplify this expression even further by
ordering the eigenvalues $x_i$ since such permutations are also
$\mathrm{U}(2s+1)$ transformation. Thus, dropping all subdominant terms, and
scaling ${\bf x}$, \bea
\mathrm{d}\mu (\phi) & \approx & \frac{1}{Z} \sum_{i=0}^{2s+1} {2s+1
  \choose i} \delta
\left(\phi-x_0 \, U^\dag \mathrm{Diag}(\mathrm{sign}(i-j+1/2)+
\sqrt{\frac{(2s+1)\lambda}{4\pi}} \frac{y_j}{rR}) \, 
U\right) \\ & & \left( \ex^{-x_0^2 K(U,\mathrm{sign}(i-j+1/2),0)}
\,\mathrm{d}U \right) \left( (\frac{16\pi r^2R^2}{\lambda(2s+1)})
^{i(2s+1-i)} \ex^{-{\bf y}\cdot {\bf y}} V^2(y_{j\leq i})V^2(y_{j>i})
\, \mathrm{d}{\bf y} \, \right) ,\label{dmu} \ena 
where each term of the sum corresponds to an integration of the action
in the vicinity of the $\mathrm{U}(2s+1)$ orbit of ${\bf 1}_i \oplus -{\bf
1}_{2s+1-i}=\mathrm{sign}(i-j+1/2)$. In this sum, the ``kinetic'' term
(containing $\mathrm{d}U$) decreases with $i$ while the ``potential''
term (containing $\mathrm{d}{\bf x}$) increases with $i$. Thus, the
matrix phase corresponds to the case when the potential term
dominates, whereas the uniform phase would correspond to one where the
kinetic term dominates. 

Note that, by reabsorbing the Vandermonde determinants into matrix
integrations, the potential term can be rewritten as a Gaussian
measure \bea
\ex^{-{\bf x}\cdot {\bf x}} V^2(x_{j\leq i})V^2(x_{j>i}) \,
\mathrm{d}{\bf x} & = & \left(\int_{\mathrm{SU}(i)} \frac{\mathrm{d}U_1}{
\mathcal{V}(\mathrm{SU}(i))}\right) V^2(x_{j\leq i}) (\prod_{j=1}^i \mathrm{d}x_j)\,
\ex^{-\sum_{j=1} ^i x_j^2} \\ 
& & \left( \int_{\mathrm{SU}(2s-i)} \frac{\mathrm{d}U_2}{\mathcal{V}(\mathrm{SU}(2s-i))}
\right) (\prod_{j=i+1}^{2s+1-i} \mathrm{d}x_j)\, \ex^{-\sum_{j=i+1}
^{2s+1-i} x_j^2} \\
& \propto & \mathcal{V}(\mathrm{Gr}_{i,2s+1})\, \delta(\psi_1-U_1^\dag
\mathrm{Diag}(x_{j\leq i})\, U_1)\, \delta(\psi_2-U_2^\dag \mathrm{Diag}
(x_{j>i})\, U_2) \\ & & \int_{\mathrm{SU}(i)\times \mathrm{SU}(2s-i)} \mathrm{d}\psi_1
\mathrm{d}\psi_2 \, \ex^{-\Tr(\psi_1^2+\psi_2^2)} ,\ena
where $\mathcal{V}$ denotes the volume of a space. This Gaussian
integral can be calculated exactly for most expectation
values. Furthermore, it makes explicit the volumes of the orbits
$\mathcal{O}_i\sim \mathrm{Gr}_{i,2s+1}$ which weights these
integrals. Not surprisingly now, the power $i(n-i)$ which appears in
the potential term is precisely the dimension of this Grassmanian.

Note also that for the purely potential action, the terms $i=s$ or
$i=s\pm 1/2$, whichever are integer, are always dominant over the
others, and so in this limit (which excludes the disordered phase
because of the approximation (\ref{expanphi})) only the matrix phase
can arise.

This extra phase does not appear for the scalar field simulated on the
lattice despite the fact that the action has the same superficial
properties. The potential has a large subset of minima given by
$\phi_{ij}=\pm x_0$ at each lattice site. The minimum with the largest
phase space is the one with half their sites with value $+x_0$ and
half with the opposite value $-x_0$ which has a degeneracy of ${N^2
\choose N^2/2}$ where $N^2$ is the number of lattice points, whereas
the minimum of the action is the one with a uniform value at each site
which has just degeneracy two. However, on the lattice the kinetic
term simply cannot be neglected because it represents the only, and
thus {\em dominant}, coupling between the degrees of freedom at each
lattice point. Thus, it must lift the degeneracy of the minimum of the
potential in favour of the usual uniform phase. By contrast,
on the fuzzy sphere, the non--locality of the potential implies that
the coupling of the matrix degrees of freedom is ensured by both the
kinetic and potential term.
\paragraph{The equilibrium configuration} As described above, in the matrix
phase, the field will settle in the vicinity of the orbit
$\mathcal{O}_s$. However, although this orbit is degenerate with
respect to the potential term (\ref{potV}), it is not with respect to
the full action because the kinetic term will lift this degeneracy.

Taking into account the kinetic term, the most probable configuration
of the field will be given by the minimum of the kinetic term (or
equivalently of the action) on the orbit $\mathcal{O}_s$. The theory
described by the scalar action restricted to the orbits
$\mathcal{O}_s$ has been studied in detail in \cite{Sachin}. The
configurations found to minimise the action on these orbits
$\mathcal{O}_i$ were found to be of the form $x_0 W^\dag \, ({\bf 1}_i
\oplus -{\bf 1}_{2s+1-i})\, W$, with $W\in SU_s (2)$ the $2s+1$
dimensional representation of $\mathrm{SU}(2)$. So, the equilibrium
configuration in the matric phase is found to be of the form $\pm x_0
W^\dag \, ({\bf 1}_s \oplus -{\bf 1}_{s+1}) W$, with $W\in SU_s (2)$.

\begin{figure} \input{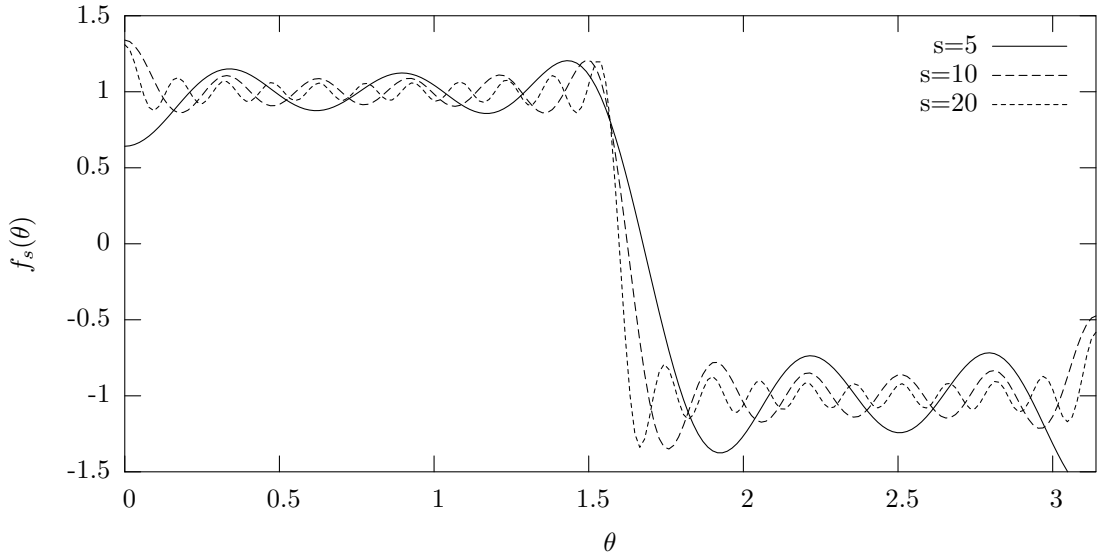} \caption{The function $f_s(\theta)$
    defined in Eq. (\ref{fs}) for various values of the matrix size
    $2s+1$. Note how the functions tend to a sign
    function.}\label{maps} \end{figure}

Applying the mapping (\ref{map}), it is possible to extrapolate from
there the form this field will take in the commutative limit. Since this
mapping preserves the action of $\mathrm{SU}(2)$, the corresponding set of
fields will be given by $\pm x_0 W^\dag \, \mathcal{M}_s ({\bf 1}_s
\oplus -{\bf 1}_{s+1})\, W$, where $W\in SO(3)$ is now a simple global
rotation. Figure \ref{maps} shows the function with azimuthal symmetry \be
f_s(\theta)=\mathcal{M}_s (-{\bf 1}_{s} \oplus {\bf 1}_{s+1})
,\label{fs} \en
with $\theta$ the zenith angle.  Thus the matrix phase on the fuzzy
sphere has a commutative limit similar to the one observed on the torus
\cite{ftorus}. It also suggests {\it a posteriori} that the name of
``striped phase'' might also be appropriate for this phase.

\begin{figure} \input{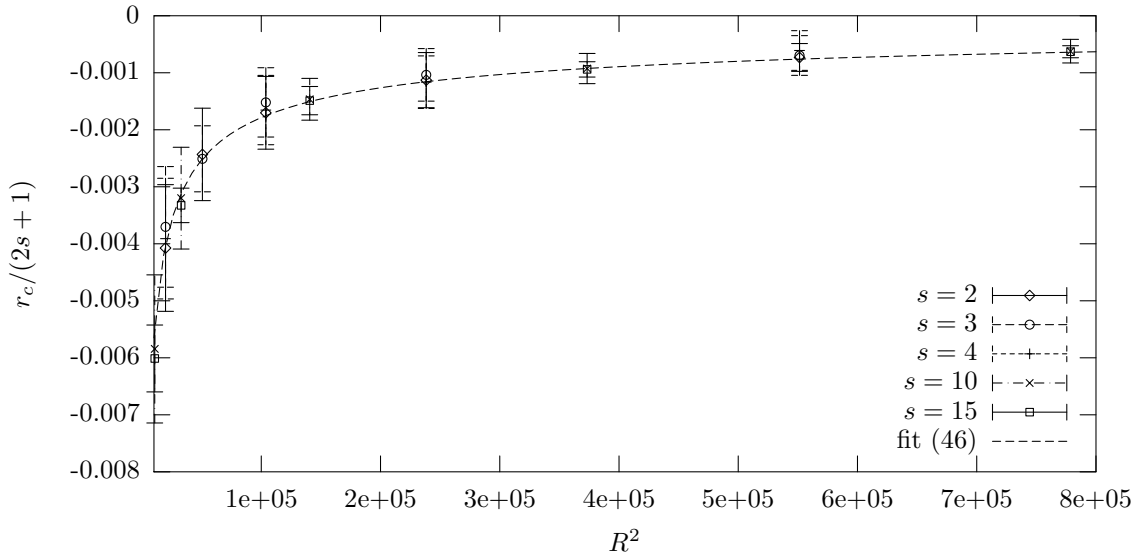}
\caption{Scaled phase diagram for the disordered--matrix phase
 transition for five distinct matrix sizes $(2s+1)\times(2s+1)$ and the
 asymptotic limit (\ref{fit2}) which fits the curves well for $R^2\geq
 30$.}
\label{phase2} \end{figure}

\paragraph{The critical line} The critical line for this phase
transition scales linearly with the matrix size, i.e. like $s$. Figure
\ref{phase2} shows the critical lines. For large radii, a fit can
easily be deduced algebraically from the collapse of the
susceptibility curves and the critical point for the pure potential
model shown in Figure \ref{compareff}a, and the scaling of the
susceptibility for the pure potential model shown in
(\ref{scaleV}). Indeed, for $s=15$, and $R\rightarrow +\infty$ \be
r^{d-m}_c(R^2) \sim \frac{-17.5}{R},\en
thus, putting back in the linear scaling found for the critical
line, \be 
\frac{r^{d-m}_c(R^2)}{2s+1} \sim \frac{-17.5}{31R}\simeq -0.56 \,
(R^2)^{-1/2} ,\label{fit2} \en 
which fits the critical line quite well as shown in Figure \ref{phase2}.

\subsection{The matrix--uniform phase transition}
This transition is difficult to observe numerically because of
thermalisation problems. To switch from the matrix phase to the
disordered phase requires a large change in the field between two
local minima of the action, which has numerically vanishing low
probability of happening. In fact, starting from hot initial conditions
(i.e. random), this transition never appears and the field stays
trapped in the matrix phase. This is quite understandable since we
have seen in Eq. (\ref{dmu}) that the phase space asociated with the
matrix phase is so much bigger than the one asociated with the uniform
phase.

Conversely, with cold initial conditions (i.e. the field is a minimum
of the action or equivalently in the uniform phase), when $-r$ gets
large enough, the field stays trapped near this locally stable
configuration, and a matrix--uniform phase appears. However, since hot
initial conditions give a different result, the observed
transition just pinpoints the parameter $r$ when the inverse of the
probability of tunneling to the matrix phase becomes larger than the
thermalisation time of the Monte--Carlo simulation.

This is why the uniform--disordered phase diagram \ref{phase1} is
truncated for $R^2>64$. Beyond this point, the three phases uniform,
matrix, and disordered, start mixing and runs with the cold initial
conditions show the emergence of an uniform phase while the hot
do not.

Still, it is possible to guess a few things about the large $-r$
region. As discussed in subsection \ref{dmphase}, the term $i=s$ in
the sum of Eq. (\ref{dmu}) corresponds to the matrix phase while the
term $i=0$ corresponds to the uniform phase. When $R$ and $s$ are
fixed and $r$ increases, the potential term grows polynomially whereas
the kinetic one is suppressed exponentially. Thus, the uniform phase
must dominate when $-r$ becomes large enough, and there {\em is} a
matrix--uniform phase transition.

Furthermore, when the kinetic term start dominating over the potential
term in (\ref{dmu}), its exponential dependence should ensure that the
term $i=0$ in the sum quickly becomes dominant. Thus, it is unlikely
that there be other intermediate phases, asociated with other minima
of orbits $\mathcal{O}_i$ $i\not\in \{ 0,s\}$, between the matrix and
uniform phases.

Note that the pure potential model is no help here as it can only have
two phases: the disordered and matrix phases. This is not surprising
as the kinetic term is a key component in this region of the phase
diagram.

A systematic analysis of this phase transition, and by extension of
the triple point where the three phases coexist, will be presented in
the future paper \cite{fergar} already mentionned at the end of
Section \ref{disuni}.

\section{Conclusions} \label{Conclu}
In conclusion, the scalar field action on the fuzzy sphere shows the
emergence of a new phase and exhibits a new aspect of the phenomenon
commonly called $UV$--$IR$ mixing. The critical line for the
uniform--disordered phase transition is identified and scales like the
square root of the matrix dimension. The critical line for the
matrix--disordered phase transition is also identified and found to
scale linearly with the matrix size and grows linearly for large
sphere radii. Finally, the existence of a matrix--uniform phase
transition is ascertained algebraically, but could not be identified
numerically due to thermalisation problems. We also conjecture that
there is no other intermediate phase between the matrix and uniform
phases.

The matrix phase was also studied in detail, showing that its
emergence over the uniform phase is linked to the dominance of the
phase space volume of the largest orbit minimising the potential, over
the kinetic term. As a result, the matrix phase must be well
approximated for large sphere radii by a pure potential model,
i.e. one with no kinetic term. This was confirmed numerically.

Since, for a scalar field theory, the geometry of a fuzzy space is
determined by the choice of its Laplacian or equivalently of the
kinetic term of the scalar field action, the matrix phase appears to
be largely independant of the geometry. Thus, if it appears in other
scalar matrix models, it must have similar properties to those
shown here, independantly of the geometry or dimension of the
commutative limiting space. In particular, it should be possible to
verify this by comparing the results found for the fuzzy sphere and
the non--commutative lattice \cite{ftorus}.

Finally, the equilibrium configuration was identified indirectly and
shown to have a ``striped'' structure similar to what has been
observed on the non--commutative lattice. 

This is of course a verification that the naive scalar field action
studied in this paper cannot be used as an approximation of the
scalar field theory on the usual sphere. For the latter, one needs to
look at a more complicated action such as the one proposed in
 \cite{den} which adds an extra damping term proportional to
$\Tr((\mathcal{L}^2\phi)^2)$ to the action to suppress $UV$--$IR$
mixing. This is consistent with the analysis presented here as such a
term will reinforce the influence of the kinetic term and thus
suppress the matrix phase.

\section*{Acknowledgements}
I wish to thank in no particular order D. O'Connor, B. Dolan,
J. Aguilar, and F. Garc\'\i a for some useful discussions on the
nature of the matrix phase, and O. Jahn, W. Bietenholz, F. Hofheinz,
and J. Nishimura for some assistance in understanding the numerical
approach to this problem.

This research was supported through a European Community Marie Curie
Fellowship. It also received partial support from the EU Research
Training Network in Quantum Spaces--Noncommutative Geometry QSNG.

\end{document}